\documentclass[12pt]{article}
\usepackage{epsfig,amsmath,amssymb,mathrsfs}

  {\end{list}}%

\makeatletter
\renewcommand{\section}{\setcounter{equation}{0}\@startsection
 {section}%
 {1}%
 {0pt}%
 {-1\baselineskip}%
 {0.4\baselineskip}%
 {\bfseries\large}}%
\renewcommand{\subsection}{\@startsection
 {subsection}%
 {2}%
 {0pt}%
 {-0.75\baselineskip}%
 {0.2\baselineskip}%
 {\bfseries}}%
\renewcommand{\subsubsection}{\@startsection
 {subsubsection}%
 {3}%
 {0pt}%
 {-0.5\baselineskip}%
 {0.1\baselineskip}%
 {\sc}}%
\makeatother

\DeclareMathAlphabet{\mathpzc}{OT1}{pzc}{m}{it}

\def\a{\alpha}
\def\dalpha{\dot{\alpha}}
\def\btheta{\bar{\theta}}

\def\g5{\gamma_{5}}

\def\idx{\int\!\! d^4\!x}

\newcommand{\bea}{\begin{eqnarray}}
\newcommand{\eea}{\end{eqnarray}}
\newcommand{\beann}{\begin{eqnarray*}}
\newcommand{\eeann}{\end{eqnarray*}}
\newcommand{\ba}{\begin{array}}
\newcommand{\ea}{\end{array}}

 \def\g {\gamma}

\begin{document}
 \begin{titlepage}
\rightline{FTI/UCM 133-2013}
\vglue 33pt

\begin{center}

{\Large \bf The Minimal and the New Minimal Supersymmetric
Grand Unified Theories on Noncommutative Space-time}\\
\vskip 1.0true cm
{\rm C. P. Mart\'{\i}n}\footnote{E-mail: carmelo@elbereth.fis.ucm.es}
\vskip 0.1 true cm
{\it Departamento de F\'{\i}sica Te\'orica I,
Facultad de Ciencias F\'{\i}sicas\\
Universidad Complutense de Madrid,
 28040 Madrid, Spain}\\

\end{center}

{\leftskip=50pt \rightskip=50pt \noindent
We construct noncommutative versions of  both the minimal and the new minimal supersymmetric
Grand Unified Theories. The enveloping-algebra formalism is used to carry out
such constructions. The beautiful formulation of the Higgs sector of these
noncommutative theories is a consequence of fact that, in the GUTs at hand,  the ordinary Higgs
fields can be realized as elements of the Clifford algebra
$\mathbb{C}\rm{l}_{10}(\mathbb{C})$. In the noncommutative supersymmetric GUTs
we formulate, supersymmetry is linearly realized by the noncommutative fields;
but it is not realized by the  ordinary fields that define those noncommutative fields via the
Seiberg-Witten map.

\par}

\vspace{9pt}
\vfill
\end{titlepage}

\section{Introduction}

Let us begin by saying that  by canonical noncommutative space-time --or simply noncommutative space-time-- we mean the noncommutative space defined by
$[X^{\mu},X^{\nu}]=i\omega^{\mu\nu}$, where $\omega^{\mu\nu}$ is a c-number. We shall assume that Lorentz indices are raised and lowered
with the Minkowski metric $(-,+,+,+)$.

The formulation of gauge theories on canonical noncommutative space-time that are deformations of
ordinary gauge theories for arbitrary gauge groups in arbitrary unitary representations demands, as yet,
using the enveloping-algebra formalism. This formalism was set up in Refs.~\cite{Madore:2000en, Jurco:2000ja,
  Jurco:2001rq} and put to use in the construction of the noncommutative Standard Model~\cite{Calmet:2001na}, a noncommutative
deformation of the ordinary Standard Model with no new degrees of freedom --see Refs.~\cite{Chaichian:2001py, Khoze:2004zc, Arai:2006ya} for
other noncommutative extensions of the ordinary Standard Model. The enveloping-algebra formalism was also employed~\cite{Aschieri:2002mc} to formulate GUTs in
the SU(5) and SO(10) gauge group cases. The nontrivial issue of constructing Yukawa terms --for SO(10) and $\rm{E}_6$-- within the enveloping-algebra framework was tackled in
Ref.~\cite{Martin:2010ng}. Outside the enveloping-algebra formalism, the formulation of noncommutative gauge theories for SO(N) groups was also discussed in Ref.~\cite{Bonora:2000td}.

In the enveloping-algebra formalism the noncommutative gauge fields belong to the universal enveloping algebra of the Lie algebra of the ordinary gauge group and they are defined in terms of the ordinary fields by means of the Seiberg-Witten map. Let us recall that the Seiberg-Witten map maps ordinary gauge orbits into noncommutative gauge orbits. When the Seiberg-Witten map is computed as a formal power series in the noncommutativity matrix parameter $\omega^{\mu\nu}$, the action of the noncommutative theory is expressed as a formal power series in $\omega^{\mu\nu}$ with coefficients that are
integrated polynomials in the ordinary fields and their derivatives.
Many theoretical properties  --e.g., renormalizability~\cite{Buric:2005xe, Buric:2006wm, Buric:2007ix, Martin:2009sg, Tamarit:2009iy, Martin:2009vg}, gauge anomalies~\cite{Martin:2002nr, Brandt:2003fx}, existence of noncommutative deformations of ordinary instantons and monopoles~\cite{Martin:2005vr, Martin:2006px, Stern:2008wi}-- of the noncommutative gauge theories so defined have been studied by taking the first few terms of the appropriate  $\omega^{\mu\nu}\!$-expanded actions. Some phenomenological properties of the noncommutative gauge theories at hand have been analyzed in~\cite{Melic:2005su, Alboteanu:2006hh, Buric:2007qx, Tamarit:2008vy, Haghighat:2010up, Trampetic:2009vy}.

The UV/IR mixing effects~\cite{Minwalla:1999px} that occurs in the $\omega^{\mu\nu}\!$-unexpanded noncommutative field theories cannot be exhibited in the noncommutative gauge theory constructed by defining the
Seiberg-Witten map as a series expansion in $\omega^{\mu\nu}\!$, unless some re-summation of an infinite number of terms in powers of  $\omega^{\mu\nu}$ is carried out: a daunting task. Fortunately, for
the enveloping-algebra formalism to work~\cite{Jurco:2001rq} it is not
necessary that the Seiberg-Witten be given by a formal series expansion in
the noncommutativity matrix
$\omega^{\mu\nu}$. Indeed, the enveloping-algebra formalism works equally well if one considers the Seiberg-Witten map as being given by an expansion in the number of ordinary fields, thus leaving its
dependence on $\omega^{\mu\nu}$ exact. Hence, if one wants to study noncommutative UV/IR effects in theories defined within the enveloping-algebra formalism one should use this $\omega^{\mu\nu}\!$-exact
Seiberg-Witten map. This was done for the first time in Ref.~\cite{Schupp:2008fs} were it was shown, in the $U(1)$ case with fermions in the adjoint, that if the $\omega^{\mu\nu}$ dependence of the Seiberg-Witten is handled exactly, then, there is an UV/IR mixing phenomenon in the
noncommutative theory defined within the enveloping-algebra formalism. The analysis of the UV/IR mixing effects was later
extended~\cite{Raasakka:2010ev} to fermions in the fundamental representation coupled to $U(1)$ gauge fields. The UV/IR mixing that occurs in the
one-loop propagator of adjoint fermions coupled to U(1) fields and its very
interesting implications on neutrino physics has been deeply analyzed in
Refs.~\cite{Horvat:2011iv, Horvat:2011bs, Horvat:2011qn, Horvat:2011qg} --see
Ref.~\cite{Trampetic:2013ey} for a  recent short review. It is worth
mentioning that the cohomological technics developed in
Refs.~\cite{Barnich:2001mc, Barnich:2003wq} --see also~\cite{Ulker:2007fm}--
are extremely helpful~\cite{Martin:2012aw} in the computation of the ($\omega^{\mu\nu}\!$-exact) 
expansion of the Seiberg-Witten map in the number of fields.

Ordinary (i.e., on Minkowski space-time) SO(10) GUTs --see
Ref.~\cite{Senjanovic:2011zz}  for a status review-- provide appealing extensions of
the Standard Model, for the 16 spinor representation of SO(10) unifies
--within each family-- the
fermionic matter of the Standard Model plus a right-handed neutrino. This is in
addition to the unification of the interactions. They also yield tiny neutrino masses
through the see-saw mechanism. The minimal supersymmetric Grand Unified Theory~\cite{Aulakh:1982sw, Aulakh:2003kg} has also
 other nice features such as
$b-\tau$ unification and leads to realistic phenomenology if split supersymmetry
is at work~\cite{Bajc:2008dc}. Another way to iron out the problems that
the original minimal supersymmetric SO(10)  GUT gave rise to is to include in
it a Higgs in the 120 irrep of SO(10). This proposal was put forward in
Ref.~\cite{Aulakh:2005mw},  were the theory was named
the new minimal supersymmetric GUT. An extensive analysis of the new minimal supersymmetric GUT has been
presented in Ref.~\cite{Aulakh:2008sn}.

The purpose of this paper is to formulate the corresponding counterparts of the
minimal supersymmetric  and the new minimal supersymmetric GUTs, which we
have just mentioned, on canonical noncommutative space-time. Two preliminary
comments are in order. First, these GUTs are particularly adequate for their generalization to noncommutative space-time, for all the Higgs fields in them have --see Sec. 3-- a beautiful
interpretation as appropriate elements of  the Clifford algebra  $\mathbb{C}\rm{l}_{10}(\mathbb{C})$; and recall that 
associative unital algebras are key mathematical objects in noncommutative
geometry~\cite{GraciaBondia:2001tr}. This is a feature not shared with SO(10)
GUTs carrying Higgs fields in the 16, 54, etc... irreps of SO(10)~\cite{Babu:1993we, Lucas:1995ic}, which nonetheless
should admit noncommutative versions within the enveloping-algebra
formalism. Second, it is known~\cite{Seiberg:1999vs,Martin:2008xa} that the
supersymmetry of the effective  U(1) supersymmetric DBI action for open strings ending on D-branes
in the presence of a constant Neveu-Schwarz $B_{\mu\nu}$ field is a nonlinearly realized supersymmetry when
the DBI action is written in terms of the ordinary gauge field and its
superpartners; whereas is a linearly realized supersymmetry when that action
is expressed, upon using the Seiberg-Witten map, in terms of corresponding
noncommutative fields. Hence, when  formulated in
terms of ordinary fields, the supersymmetry of the noncommutative U(1) theory is not
the supersymmetry  of the corresponding ordinary theory, which is obtained by setting the
noncommutativity parameter to zero. It also happens~\cite{Martin:2008xa} that
noncommutative U(N) superYang-Mills
has a linearly realized supersymmetry if the theory is expressed in terms of
noncommutative fields,  and yet that supersymmetry has a nonlinear realization when,
upon using the Seiberg-Witten map, ordinary fields are chosen to formulate the
theory. If we have SU(N),   the supersymmetric invariance of the noncommutative
supersymmetric theory is linearly realized in terms of the noncommutative
fields, but cannot be  realized by using the ordinary fields that define the
former noncommutative fields via the Seiberg-Witten map --we shall see that
this very situation occurs for the noncommutative GUTs that we shall construct. Let us also
mention that in the SU(N) case the one-loop UV divergent radiative corrections preserve, up to first order in $\omega^{\mu\nu}\!$, the structure of classical action that is consistent with having  linearly realized supersymmetry when the action is expressed in terms of the noncommutative fields --see~\cite{Martin:2009mu} for details. It would thus appear that some nice properties of
ordinary supersymmetric theories are still maintained through its --although hidden--  noncommutative linear realization.

The layout of this paper is as follows. In Sec. 2, we discuss how to obtain
the field content and action of the noncommutative minimal supersymmetric
Grand Unified Theory from the  noncommutative new minimal supersymmetric
Grand Unified Theory. Sec. 3 is a  summary of the field content and
action of the new minimal supersymmetric Grand Unified Theory on ordinary
Minkowski space-time.  We  formulate the theory in terms of ordinary superfields
in the Wess-Zumino gauge and interpret its Higgs superfields as elements of
$\mathbb{C}\rm{l}_{10}(\mathbb{C})$, for this is most suitable for  its noncommutative
generalization. Sec. 4 is devoted to the  construction of our
noncommutative counterpart of the new minimal supersymmetric Grand Unified
Theory by using the enveloping-algebra formalism. In Sec. 5 of the
paper, we make some comments on the fact that in  the noncommutative theory
formulated in the previous section supersymmetry, which is linearly realized by
the noncommutative fields,  is not realized by the corresponding ordinary fields.

\section{The Noncommutative Minimal Supersymmetric Grand Unified Theory}

The action of the noncommutative minimal supersymmetric Grand Unified
Theory is obtained from the action of the noncommutative new minimal supersymmetric Grand
Unified Theory by removing from the latter  the Higgs superfield that is constructed from the ordinary Higgs field 
transforming under the 120 irrep of SO(1O). Hence we shall move on directly
to the construction of the noncommutative new minimal supersymmetric Grand Unified Theory.

\section{The New Minimal supersymmetric Grand Unified Theory on Minkowski space-time}

The new minimal supersymmetric Grand Unified Theory was introduced in
Ref.~\cite{Aulakh:2005mw} -- see also Ref.~\cite{Aulakh:2008sn}. Let us spell out its superfield
content.  First, three --one for each
family in the Standard Model-- chiral scalar superfields,
$\Phi^{(16)}_{f},\,f=1,2,3$, transforming under the 16 irrep of SO(10). The
$\Phi^{(16)}_{f}$'s contain the
fermion fields of the Standard Model plus a right-handed neutrino. Secondly, five Higgs
chiral scalar superfields, $\Phi^{(210)}_{i_1i_2i_3i_4}, \Phi^{(10)}_{i_1},
\Phi^{(126)}_{i_1i_2i_3i_4i_5}, \Phi^{(\overline{126})}_{i_1i_2i_3i_4i_5}$,
and $\Phi^{(120)}_{i_1i_2i_3}$ transforming, respectively, under the 210, the
10, the 126, the $\overline{126}$ and the 120  irreps of SO(10). The indices $i_1,i_2,....$ run
from 1 to 10,  and  $\Phi^{(210)}_{i_1i_2i_3i_4}$,
$\Phi^{(126)}_{i_1i_2i_3i_4i_5}$, $\Phi^{(\overline{126})}_{i_1i_2i_3i_4i_5}$
and $\Phi^{(120)}_{i_1i_2i_3}$ are totally antisymmetric SO(10) tensors with
regard to its  $i_1, i_2,..$ indices. Further,
$\Phi^{(126)}_{i_1i_2i_3i_4i_5}$, $\Phi^{(\overline{126})}_{i_1i_2i_3i_4i_5}$
satisfy the following  duality equations:
\begin{equation*}
\begin{array}{c}
{\Phi^{(126)}_{i_1i_2i_3i_4i_5}=-\frac{i}{5!}\,\varepsilon_{i_1i_2i_3i_4i_5i_6i_7i_8i_9i_{10}}\,\Phi^{(126)}_{i_6i_7i_8i_9i_{10}}},\\[4pt]
{\Phi^{(\overline{126})}_{i_1i_2i_3i_4i_5}=+\frac{i}{5!}\,\varepsilon_{i_1i_2i_3i_4i_5i_6i_7i_8i_9i_{10}}\,\Phi^{(\overline{126})}_{i_6i_7i_8i_9i_{10}}}.
\end{array}
\end{equation*}
Finally, there is the vector superfield, $V$, taking values in the appropriate --see below-- 
representation of SO(10). In the Wess-Zumino gauge, $V$  reads
\begin{equation*}
V=-\theta\sigma^{\mu}\bar{\theta}\,a_{\mu}+i\theta^{2}\bar{\theta}\lambda-i\bar{\theta}^{2}\theta\lambda+\frac{1}{2}
\theta^2\bar{\theta}^2D.
\end{equation*}
 Here we shall adopt the supersymmetry conventions of Ref.~\cite{Wess:1992cp}.

Let $\Gamma^{i}$ denote the Dirac matrices in 10 Euclidean dimensions. These
matrices generate the Clifford algebra $\mathbb{C}\rm{l}_{10}(\mathbb{C})$.
We shall see later that a  noncommutative version of the
new minimal supersymmetric Grand Unified Theory can be constructed in a very
smart way by using the $\mathbb{C}\rm{l}_{10}(\mathbb{C})$ Clifford algebra valued Higgs superfields
\begin{equation}
\begin{array}{l}
{\Phi^{(210)}=\Gamma^{i_1}\Gamma^{i_2}\Gamma^{i_3}\Gamma^{i_4}\Phi^{(210)}_{i_1i_2i_3i_4},\;
\Phi^{(10)}=\Gamma^{i_1}\Phi^{(10)}_{i_1},\;
\Phi^{(126)}=\Gamma^{i_1}\Gamma^{i_2}\Gamma^{i_3}\Gamma^{i_4}\Gamma^{i_5}\Phi^{(126)}_{i_1i_2i_3i_4i_5},}\\[4pt]

{\Phi^{(\overline{126})}=\Gamma^{i_1}\Gamma^{i_2}\Gamma^{i_3}\Gamma^{i_4}\Gamma^{i_5}\Phi^{(\overline{126})}_{i_1i_2i_3i_4i_5},\;
\Phi^{(120)}=\Gamma^{i_1}\Gamma^{i_2}\Gamma^{i_3}\Phi^{(120)}_{i_1i_2i_3}}.
\end{array}
\label{Clifhiggs}
\end{equation}
rather than the SO(10) tensor superfields
$\Phi^{(210)}_{i_1i_2i_3i_4}$, $\Phi^{(10)}_{i_1}$, $\Phi^{(126)}_{i_1i_2i_3i_4i_5}$,
$\Phi^{(\overline{126})}_{i_1i_2i_3i_4i_5}$ and $\Phi^{(120)}_{i_1i_2i_3i_4}$, 
which give rise to the former.

From now on, the symbol $V$ will stand for the vector superfield in the
Wess-Zumino gauge whose supersymmetric components take values in the
$16\bigoplus\overline{16}$ representation of SO(10):
\begin{equation}
\begin{array}{l}
{V=\frac{1}{2}\Sigma^{ij}V^{ij},\quad\Sigma^{ij}=\frac{1}{4i}\,[\Sigma^i,\Sigma^j],\quad
  i,j=1...10,}\\[4pt]
{V^{ij}=-\theta\sigma^{\mu}\bar{\theta}\,a^{ij}_{\mu}+i\theta^{2}\bar{\theta}\bar{\lambda}^{ij}-i\bar{\theta}^{2}\theta\lambda^{ij}+\frac{1}{2}
\theta^2\bar{\theta}^2 D^{ij}.}
\end{array}
\label{ordsvec}
\end{equation}
$V^{ij}$ carry the 45 irrep of SO(10). Below, we shall use the notation
\begin{equation}
a_{\mu}=\frac{1}{2}\Sigma^{ij}a^{ij}_\mu,\,
\lambda^{ij}=\frac{1}{2}\Sigma^{ij}\lambda^{ij},\, \bar{\lambda}^{ij}=\frac{1}{2}\Sigma^{ij}\bar{\lambda}^{ij},\,
D=\frac{1}{2}\Sigma^{ij}D^{ij}.
\label{ordsmult}
\end{equation}

Let us introduce now the chiral coordinate $y=x+i\theta\sigma^{\mu}\bar{\theta}$.
Let $\Lambda$ be the chiral superfield defined as follows
\begin{equation*}
\begin{array}{l}
{\Lambda=\frac{1}{2}\Lambda^{ij}\Sigma^{ij},}\\[4pt]
{\Lambda^{ij}(y)=-2i\theta\sigma^{\mu}\bar{\xi} a^{ij}_{\mu}(y)- 2\theta^2
\bar{\xi}\bar{\lambda}^{ij}(y),}
\end{array}
\end{equation*}
where $\bar{\xi}$ is an infinitesimal spinor. Then, the supersymmetry
transformation of the vector superfield we have introduced
--recall that we have chosen the Wess-Zumino gauge-- reads
\begin{equation}
\delta^{WZ}_{\xi}V\,=\,(\xi Q+\bar{\xi}\bar{Q})\,V\,+\,
\delta_{\Lambda}\,V,
\label{ordsusytrans}
\end{equation}
where
\begin{equation}
Q_{\alpha}=\frac{\partial}{\partial\theta^{\alpha}}-i\sigma^{\mu}_{\alpha\dalpha}\btheta^{\dalpha}\partial_{\mu},\;
\bar{Q}_{\dalpha}=-\frac{\partial}{\partial\theta^{\dalpha}}+i\theta^{\alpha}\sigma^{\mu}_{\alpha\dalpha}\partial_{\mu}
\label{theQs}
\end{equation}
and $\delta_{\Lambda}\,V$ is given by the following compensating gauge
transformation:
\begin{equation}
\delta_{\Lambda}\,V=\frac{i}{2}{\cal
  L}_{V}(\Lambda+\bar{\Lambda})+\frac{i}{2}{\cal L}_{V}
\coth {\cal
  L}_{V}(\Lambda-\bar{\Lambda}),\quad {\cal L}_{V}\,F= [V, F].
\label{compentrans}
\end{equation}
In the Wess-Zumino gauge, the supersymmetry transformation of the scalar
superfield $\Phi^{(16)}_f$ reads
\begin{equation*}
\delta^{WZ}_{\xi}\Phi^{(16)}_f\,=\,(\xi\ Q+\bar{\xi}\bar{Q})\,\Phi^{(16)}_f\,+\,
\delta_{\Lambda}\,\Phi^{(16)}_f,\quad \delta_{\Lambda}\,
\Phi^{(16)}_f=-i\,\Lambda\Phi^{(16)}_f.
\end{equation*}

And last, but not least, the $\mathbb{C}\rm{l}_{10}(\mathbb{C})$ Clifford algebra valued Higgs
superfields in~(\ref{Clifhiggs}) transform under supersymmetry in the
Wess-Zumino gauge as follows
\begin{equation*}
\delta^{WZ}_{\xi}\Phi^{(H)}\,=\,(\xi Q+\bar{\xi}\bar{Q})\,\Phi^{(H)}\,+\,
\delta_{\Lambda}\,\Phi^{(H)},\quad \delta_{\Lambda}\,
\Phi^{(H)}=-i\,[\Lambda,\Phi^{(H)}],
\end{equation*}
where $\Phi^{(H)}$ stands for any of the scalar superfields defined in~(\ref{Clifhiggs}).

The superfields $V,$$\Phi^{(16)}_f,$$\Phi^{(210)},$$\Phi^{(10)},$
$\Phi^{(126)},$$\Phi^{(\overline{126})}$ and $\Phi^{(120)}$ give a
redundant characterization of the physical system, for there is
still the invariance under the following gauge transformation
\begin{equation*}
\begin{array}{l}
{\delta_{\Omega}V=\frac{i}{2}{\cal
  L}_{V}(\Omega+\bar{\Omega})+\frac{i}{2}{\cal L}_{V}
\coth {\cal
  L}_{V}(\Omega-\bar{\Omega}),}\\[4pt]
{\delta_{\Omega}\Phi^{(16)}_f = -i, \Omega\, \Phi^{(16)}_f,\,
  \delta_{\Omega}\Phi^{(H)} = -i, [\Omega,\,\Phi^{(H)}]},\quad H=210, 10, 126,
\overline{126}, 120.
\end{array}
\end{equation*}
Notice that $\Omega=\frac{1}{2}\Omega^{ij}(y)\Sigma^{ij}$, $\Omega^{ij}(x)$ being
infinitesimal real functions.

Let us define the action, $S$, of the new minimal Grand Unified Theory in
terms of the superfields introduced above:
\begin{equation*}
S\,=\,S_{YM}\,+\,S_{V\Phi}\,+\,S_{spot},
\end{equation*}
where
\begin{equation}
\begin{array}{l}
{S_{SYM}=\frac{1}{64\pi}{\rm Im}\Big\{\tau\idx\,d^2\theta\,W^\alpha
  W_{\alpha}\Big\},\; \tau=\frac{\theta_{YM}}{2\pi}+\frac{4\pi i}{g^2},}\\[4pt]
{S_{V\Phi}\,=\,\idx\,d^2\theta\,d^2\bar{\theta}\;
 \sum_{f}\,(\Phi^{(16)})^{\dagger}_f\,e^{2V}\,\Phi^{(16)}_f\,+\,\sum_{H}\,\frac{1}{s(H)}{\rm
   Tr}\Big((\Phi^{(H)})^{\dagger}\,e^{2V}\Phi^{(H)}e^{-2V}\Big),}\\[4pt]
{S_{spot}=\idx\,d^2\theta\;\{ W_{matter}\,+\,W_{Higgs}\}\,+\,{\rm h.c.},}
\end{array}
\label{sactions}
\end{equation}
with
\begin{equation*}
W_\alpha=-\frac{1}{4}\,\bar{D}^2(e^{-2V}D_\alpha e^{2V})
\end{equation*}
and with $H$ running over the 210, the 10, the 126,  the $\overline{126}$ and the
120 irreps of SO(10). In~(\ref{sactions}),  the coefficients $s(H)$ are symmetry factors with values
$s(210)=1/32(1/4!)^2$,  $s(10)=1/32$, $s(126)=-1/64(1/5!)^2$,
$s(\overline{126})=-1/64(1/5!)^2$ and $s(120)=-(1/3!)^2$.  $W_{matter}$
and $W_{Higgs}$ in~(\ref{sactions}) denote the superpotentials, which read
\begin{equation}
\begin{array}{l}
{ W_{matter}=\sum_{f,f'}\;\Big\{{\cal
      Y}^{(10)}_{ff'}\;\tilde{\Phi}^{(16)}_f\Phi^{(10)} \Phi^{(16)}_{f'}}\\[4pt]
{\phantom{W_{matter}=}+{\cal
      Y}^{(\overline{126})}_{ff'}\;\tilde{\Phi}^{(16)}_f\Phi^{(\overline{126})}\Phi^{(16)}_{f'}+

{\cal
      Y}^{(120)}_{ff'}\;\tilde{\Phi}^{(16)}_f \Phi^{(120)}\Phi^{(16)}_{f'}\Big\}}\\[4pt]
{\text{and}}\\[4pt]
{W_{Higgs}= \frac{M^{(210)}}{64 (4!)^2}{\rm Tr}\,\Phi^{(210)}\Phi^{(210)}
           -\frac{M^{(126)}}{32 (5!)^2}{\rm Tr}\,\Phi^{(\overline{126})}\Phi^{(126)}
           + \frac{M^{(10)}}{64 }{\rm Tr}\,\Phi^{(10)}\Phi^{(10)}}\\[4pt]
{\phantom{W_{Higgs}=}
           -\frac{M^{(120)}}{64 (3!)^2}{\rm
             Tr}\,\Phi^{(120)}\Phi^{(120)}}\\[4pt]
{\phantom{W_{Higgs}=}+\lambda_1\,{\rm Tr}\,\Phi^{(210)}\Phi^{(210)}\Phi^{(210)}+\lambda_2 {\rm Tr}\,\Phi^{(210)}\Phi^{(\overline{126})}\Phi^{(126)}+\lambda_3 {\rm Tr}\,\Phi^{(10)}\Phi^{(120)}\Phi^{(210)}}\\[4pt]
{\phantom{W_{Higgs}=}+\lambda_4\,{\rm
      Tr}\,\Phi^{(120)}\Phi^{(210)}\Phi^{(126)}+\lambda_5 {\rm Tr}\,\Phi^{(10)}\Phi^{(210)}\Phi^{(126)}
+\lambda_6 {\rm Tr}\,\Phi^{(10)}\Phi^{(210)}\Phi^{(\overline{126})}}\\[4pt]
{\phantom{W_{Higgs}=}+\lambda_7\,{\rm
      Tr}\,\Phi^{(120)}\Phi^{(120)}\Phi^{(210)}+\lambda_8 {\rm Tr}\,\Phi^{(120)}\Phi^{(210)}\Phi^{(\overline{126})}.}
\end{array}
\label{ordspoten}
\end{equation}
In the superpotential $W_{matter}$, the chiral superfield $\tilde{\Phi}^{(16)}_f$, $f=1,2,3$, is defined as follows
\begin{equation*}
\tilde{\Phi}^{(16)}_f=\Phi^{(16)}_f\,B,\quad B=\prod_{i=odd}\,\Gamma^i.
\end{equation*}
The action of $\delta^{WZ}_{\xi}$ and of
$\delta_{\Omega}$ on $\tilde{\Phi}^{(16)}_f$  read
\begin{equation*}
\begin{array}{l}
{\delta^{WZ}_{\xi}\tilde{\Phi}^{(16)}_f\,=\,(\xi\ Q+\bar{\xi}\bar{Q})\,
\tilde{\Phi}^{(16)}_f\,+\,
\delta_{\Lambda}\,\tilde{\Phi}^{(16)}_f,\quad \delta_{\Lambda}\,
\tilde{\Phi}^{(16)}_f=i\,\tilde{\Phi}^{(16)}_f\,\Lambda,}\\[4pt]
{\delta_{\Omega}\,
\tilde{\Phi}^{(16)}_f=i\,\tilde{\Phi}^{(16)}_f\,\Omega,}
\end{array}
\end{equation*}
respectively.

For further reference, we shall close this section with the expansion in
supersymmetric components of the chiral scalar superfields $\Phi^{(16)}_f$, $f=1,2,3$ and $\Phi^{(H)}$, $H=210,10,126,\overline{126}, 120$,  defined above:
\begin{equation}
\begin{array}{l}
{\Phi^{(16)}_f=A^{(16)}_f(y)+\sqrt{2}\theta\psi^{(16)}_f(y)+\theta^2
  F^{(16)}_f(y),\;f=1,2\; \text{and}\; 3,}\\[4pt]
{\tilde{\Phi}^{(16)}_f=\tilde{A}^{(16)}_f(y)+\sqrt{2}\theta\tilde{\psi}^{(16)}_f(y)+\theta^2
  \tilde{F}^{(16)}_f(y),\;f=1,2\; \text{and}\; 3,}\\[4pt]
{\Phi^{(H)}=A^{(H)}(y)+\sqrt{2}\theta\psi^{(H)}(y)+\theta^2 F^{(H)}(y),\quad
  H=210,10,126,\overline{126},  120.}
\end{array}
\label{mattercomp}
\end{equation}
$A^{(16)}_f,$$\psi^{(16)}_f$ and $F^{(16)}_f$ transform under the 16
of SO(10). $\tilde{A}^{(16)}_f=A^{(16)}_f\,B,$$\tilde{\psi}^{(16)}_f=\psi^{(16)}_f\,B$ and
$\hat{F}^{(16)}_f=F^{(16)}_f\,B$.
 $A^{(H)},$$\psi^{(H)}$ and $F^{(H)}$ take values in
 Clifford algebra $\mathbb{C}\rm{l}_{10}(\mathbb{C})$ and are constructed from
the appropriate components of the corresponding  SO(10) antisymmetric tensor superfields.

\section{The Noncommutative New Minimal Supersymmetric Grand Unified Theory}

Here we shall put forward a supersymmetric noncommutative deformation of the
new minimal supersymmetric Grand Unified Theory. This will be a
noncommutative field theory on the noncommutative superspace defined by
the triplet $(X^{\mu},\theta^{\alpha},\bar{\theta}^{\dalpha})$ satisfying the
following equations:
\begin{equation}
[X^{\mu},X^{\nu}]=i\,\omega^{\mu\nu},\;
\{\theta^{\alpha},\theta^{\beta}\}=0,\;
\{\theta^{\alpha},\bar{\theta}^{\dot{\beta}}\}=0,\;
\{\bar{\theta}^{\dalpha},\bar{\theta}^{\dot{\beta}}\}=0,\:
[X^{\mu},\theta^{\alpha}]=0,\;[X^{\mu},\bar{\theta}^{\dalpha}]=0.
\label{comrel}
\end{equation}

Let $\xi^{\alpha}$ and $\bar{\xi}^{\dalpha}$ be infinitesimal Grassmann numbers.  Then, the
previous set of equations is invariant under supertranslations defined
thus
\begin{equation}
X^{'\mu}=X^{\mu}+i\theta\sigma^{\mu}\bar{\xi}-i\xi\sigma^{\mu}\bar{\theta},\;\theta^{'\alpha}=\theta^{\alpha}+\xi^{\alpha},\;
\bar{\theta}^{'\dalpha}=\bar{\theta}^{\dalpha}+\bar{\xi}^{\dalpha}.
\label{strans}
\end{equation}
Hence, one is naturally led to understand supersymmetry as realized by
superstranslations --modulo gauge transformations, if the Wess--Zumino gauge is
chosen-- of suitable  fields defined on the noncommutative superspace
introduced above. These suitable fields on our noncommutative superspace
--which we shall call noncommutative superfields-- will be obtained by taking any
ordinary superfield and promoting its components to the category of
noncommutative fields.  Thus we shall leave  unchanged the Grassmann structure
of the superfields. This is in harmony with the fact that there is
no deformation of the Grassmann algebra introduced in ~(\ref{comrel}).

\subsection{The noncommutative vector superfield and the superYang-Mills action}

Taking~(\ref{ordsvec}) as the starting point, we introduce first  the noncommutative vector superfield in the Wess-Zumino gauge, $\hat{V}$, of our
theory:
\begin{equation}
\hat{V}=-\theta\sigma^{\mu}\bar{\theta}\,\hat{a}_{\mu}+i\theta^{2}\bar{\theta}\bar{\hat{\lambda}}-i\bar{\theta}^{2}\theta\hat{\lambda}+\frac{1}{2}
\theta^2\bar{\theta}^2 \hat{D}.
\label{wzsvector}
\end{equation}
The components $\hat{a}_{\mu},\,\hat{\lambda},\, \bar{\hat{\lambda}}=\hat{\lambda}^{\dagger}$ and $\hat{D}$ are noncommutative
fields which --recall that we are dealing with a simple gauge group: SO(10)--  are to be constructed from their ordinary counterparts by
using the formalism put forward in Refs.~\cite{Madore:2000en, Jurco:2001rq,
  Aschieri:2002mc}. That is, $\hat{a}_{\mu},\,\hat{\lambda},\, \bar{\hat{\lambda}}$ and $\hat{D}$
are functions of $a_{\mu},\,\lambda,\, \bar{\lambda}$, $D$ --in~(\ref{ordsvec}) and~(\ref{ordsmult})-- and $\omega^{\mu\nu}$ that solve the
following  Seiberg-Witten map equations:
\begin{equation}
\begin{array}{l}
{\hat{s}\hat{\Omega}=s\hat{\Omega},}\\[4pt]
{\hat{s}\,\hat{a}_{\mu}=s\hat{a}_{\mu},\;
  \hat{s}\,\hat{\lambda}_{\a}=s\hat{\lambda}_{\a},\;
\hat{s}\,\bar{\hat{\lambda}}_{\dalpha}=s\bar{\hat{\lambda}}_{\dalpha},\;
  \hat{s}\hat{D}=s\hat{D}}.
\end{array}
\label{sweqgauge}
\end{equation}
The symbol $s$ denotes the ordinary BRS operator, which is defined as follows
\begin{equation*}
\begin{array}{l}
{s\Omega=-i\,\Omega\,\Omega,}\\[4pt]
{sa_{\mu}=\partial_{\mu}\Omega+i[a_{\mu},\;\Omega]=\mathcal{D}_{\mu}\Omega,\;
s\lambda_{\a}=i[\lambda_{\a},\Omega],\;
s\bar{\lambda}_{\dalpha}=i[\bar{\lambda}_{\dalpha},\Omega],\;
sD=i[D,\Omega].}
\end{array}
\end{equation*}
$\hat{s}$ denotes the noncommutative BRS operator, which acts on
the noncommutative fields thus:
\begin{equation}
\begin{array}{l}
{\hat{s}\hat{\Omega}=-i\,\hat{\Omega}\star\hat{\Omega},}\\[4pt]
{\hat{s}\hat{a}_{\mu}=\partial_{\mu}\hat{\Omega}+i[\hat{a}_{\mu},\;\hat{\Omega}]_{\star}=\mathcal{D}_{\mu}\hat{\Omega},\;
\hat{s}\hat{\lambda}_{\a}=i[\hat{\lambda}_{\a},\hat{\Omega}]_{\star},\;
\hat{s}\hat{\bar{\lambda}}_{\dalpha}=i[\hat{\bar{\lambda}}_{\dalpha},\hat{\Omega}]_{\star},\;
\hat{s}\hat{D}=i[\hat{D},\hat{\Omega}]_{\star}.}
\end{array}
\label{ncbrsop}
\end{equation}
The symbol $\star$ shows that functions are multiplied by using the Moyal product.
Let us remark that here $\Omega$ and $\hat{\Omega}$ are the Grassmann functions that define the
BRS transformations. Further, $\hat{\Omega}$ is a function of $a_{\mu}$ and
$\Omega$ --and the other ordinary fields, if that is our choice--
that solves $\hat{s}\hat{\Omega}=s\hat{\Omega}$ in~(\ref{sweqgauge}). One obtains a solution to~(\ref{sweqgauge}) by
particularizing  the general formulae in \cite{Martin:2012aw} to the case at hand.

Let us stress that  our definition of noncommutative vector superfield as a
function of the ordinary fields in the gauge supermultiplet, $(a_{\mu},
\lambda, D)$,  is quite in keeping with the fact that $(a_\mu,\lambda_{\a},D)$ and
$(a_\mu+\delta_{\Omega}\,a_{\mu},\lambda_{\a}+\delta_{\Omega}\,\lambda_{\a},D+\delta_{\Omega}\,D)$
characterize the same field configuration, when $\delta_{\Omega}\,a_{\mu},
\delta_{\Omega}\,a_{\mu}$ and $\delta_{\Omega}\,D$ denote infinitesimal gauge
transformations. Indeed, one can show that
\begin{equation}
\hat{V}[a_\mu+\delta_{\Omega}\,a_{\mu},\lambda_{\a}+\delta_{\Omega}\,\lambda_{\a},D+\delta_{\Omega}\,D]
=\hat{V}[a_\mu,\lambda_{\a},D]+\hat{\delta}_{\hat{\Omega}}\,\hat{V}[a_\mu,\lambda_{\a},D],
\label{harmony}
\end{equation}
where
\begin{equation}
\hat{\delta}_{\hat{\Omega}}\,\hat{V}=\frac{i}{2}{\hat{\cal
  L}}_{\hat{V}}(\hat{\Omega}+\bar{\hat{\Omega}})+
\frac{i}{2}\hat{{\cal L}}_{\hat{V}}
\coth \hat{{\cal
  L}}_{\hat{V}}(\hat{\Omega}-\bar{\hat{\Omega}}),\quad
\hat{{\cal L}}_{\hat{V}}\,F= [\hat{V}, F]_{\star}.
\label{supergt}
\end{equation}
In~(\ref{harmony}) and~(\ref{supergt}),  $\hat{\Omega}$ denotes the
chiral superfield which is obtained from $\hat{\Omega}(x)$ by replacing
$x^\mu$ with the chiral coordinate $y^\mu=x^\mu+i\theta\sigma^{\mu}\bar{\theta}$. $\hat{\Omega}(x)$, which is the image under the Seiberg-Witten map of
$\Omega$,  defines the noncommutative gauge transformations of
$\hat{a}_{\mu}$, $\hat{\lambda}_{\alpha}$ and $\hat{D}$:
\begin{equation*}
\hat{\delta}_{\hat{\Omega}}\,\hat{a}_\mu=\partial_{\mu}\hat{\Omega}+i[\hat{a}_{\mu},\hat{\Omega}]_{\star},\;
\hat{\delta}_{\hat{\Omega}}\,\hat{\lambda}_{\a}=i[\hat{\lambda}_{\a},\hat{\Omega}]_{\star},\;
\hat{\delta}_{\hat{\Omega}}\,\hat{D}=i[\hat{D},\hat{\Omega}]_{\star}.
\end{equation*}

A final comment regarding the superfield gauge transformation in~(\ref{supergt}). Let $\hat{\Omega}(y)$ be such that
$\hat{\Omega}(x)^{\dagger}=\hat{\Omega}(x)$, with an  $\hat{\Omega}(x)$ which
does not depend neither  on $\theta$ nor on $\bar{\theta}$. Then, for such an
$\hat{\Omega}(y)$, the transformation in~(\ref{supergt}) is the most general
gauge transformation of the vector superfield in the Wess-Zumino gauge which
gives a  vector superfield in the Wess-Zumino gauge.

Let us now define the supersymmetry transformations of  $\hat{V}$ introduced
above. It is plain that a supertranslation --see~(\ref{strans})-- acting
on $\hat{V}$ is generated by $\xi Q+\bar{\xi}\bar{Q}$, with $Q_{\a}$
and $\bar{Q}_{\dalpha}$ as given in~(\ref{theQs}). As in the ordinary case,
$(\xi Q+\bar{\xi}\bar{Q})\,\hat{V}$ contains more components than a
vector superfield in the Wess-Zumino gauge does, but, analogously to the
ordinary case, these extra components    are not
physical since they can be set to zero by an appropriate (field dependent)
noncommutative superfield gauge transformation. Hence, we define the infinitesimal
supersymmetry transformation of the noncommutative vector superfield as follows:
\begin{equation}
\hat{\delta}^{WZ}_{\xi}\,\hat{V}\,=\,(\xi Q+\bar{\xi}\bar{Q})\,\hat{V}\,+\,
\hat{\delta}_{\hat{\Lambda}}\,\hat{V},
\label{ncsusytrans}
\end{equation}
where $\hat{\Lambda}(y)$ is the chiral superfield
\begin{equation}
\hat{\Lambda}(y)=-2\,i\,\theta\sigma^{\mu}\bar{\xi}\,\hat{a}_{\mu}(y)-2\,\theta^2\bar{\xi}\bar{\hat{\lambda}}(y).
\label{hatlambda}
\end{equation}
and where the noncommutative superfield gauge transformation
$\hat{\delta}_{\hat{\Lambda}}\,\hat{V}$ is  obtained by replacing
$\hat{\Omega}$ with $\hat{\Lambda}$ in~(\ref{supergt}). Of course, that
$\hat{\delta}^{WZ}_{\xi}\,\hat{V}$ as defined in the previous equations looks
like the ordinary $\delta^{WZ}_{\xi}\,V$ in~(\ref{ordsusytrans}) and~(\ref{compentrans}) comes from the fact that we are not deforming the
Grassmann part of the superspace.

From~(\ref{ncsusytrans}) one readily deduces the action of $\hat{\delta}^{WZ}_{\xi}$ on the
components, $(\hat{a}_{\mu},\hat{\lambda}_{\alpha},\hat{D})$, of $\hat{V}$:
\begin{equation}
\begin{array}{l}
{\hat{\delta}^{WZ}_{\xi}\,\hat{a}_{\mu}=-i\,\bar{\hat{\lambda}}\bar{\sigma}_{\mu}\xi
+i\,\bar{\xi}\bar{\sigma}_{\mu}\hat{\lambda},}\\[4pt]
{\hat{\delta}^{WZ}_{\xi}\,\hat{\lambda}_\a = (\sigma^{\mu\nu}\xi)_\a\,\hat{f}_{\mu\nu}+i\xi_\a\,\hat{D},}\\[4pt]
{\hat{\delta}^{WZ}_{\xi}\,\hat{D} = -\xi\sigma^{\mu}\mathcal{D}_{\mu}\bar{\hat{\lambda}}-\mathcal{D}_{\mu}\hat{\lambda}\sigma^{\mu}\bar{\xi}},
\end{array}
\label{susycomp}
\end{equation}
where $\hat{f}_{\mu\nu}=\partial_{\mu}\hat{a}_{\nu}-\partial_{\nu}\hat{a}_{\mu}+i[\hat{a}_{\mu},\hat{a}_{\nu}]_{\star}$
and
$\mathcal{D}_{\mu}\hat{\lambda}_{\alpha}=\partial_{\mu}\hat{\lambda}_{\alpha}+i[\hat{a}_{\mu},\hat{\lambda}_{\alpha}]_{\star}$. It
is worth mentioning that $\hat{\delta}^{WZ}_{\xi}\,\hat{a}_{\mu}$,
$\hat{\delta}^{WZ}_{\xi}\,\hat{\lambda}_\a$ and
$\hat{\delta}^{WZ}_{\xi}\,\hat{D}$ are well-defined functions of the
infinitesimal gauge orbit of $(a_{\mu},\lambda_{\alpha},D)$, for
\begin{equation}
\hat{\delta}^{WZ}_{\xi}\,\hat{\cal X}[a_\mu+\delta_{\Omega}\,a_{\mu},
\lambda_{\a}+\delta_{\Omega}\,\lambda_{\a},D+\delta_{\Omega}\,D]=
i[\hat{\delta}^{WZ}_{\xi}\,\hat{\cal X},\hat{\Omega}],
\label{deltaclass}
\end{equation}
where $\hat{\cal X}=\hat{a}_{\mu},\,\hat{\lambda}_\a,\,D$, and
$\delta_{\Omega}$ generates an infinitesimal ordinary gauge transformation.

It can be seen that if $\hat{a}_{\mu}$, $\hat{\lambda}_{\a}$, $\hat{D}$ are
solutions to the equations~(\ref{sweqgauge}), then
\begin{equation}
\hat{a}'_{\mu}=\hat{a}_{\mu}+\hat{\delta}^{WZ}_{\xi}\,\hat{a}_{\mu},\quad
\hat{\lambda}'_\a=\hat{\lambda}_\a+\hat{\delta}^{WZ}_{\xi}\,\hat{\lambda}_\a,
\quad \hat{D}'=\hat{D}+\hat{\delta}^{WZ}_{\xi}\,\hat{D}
\label{compprime}
\end{equation}
are also solutions  to  the equations~(\ref{sweqgauge}), satisfying the  conditions
\begin{equation*}
\hat{a}'_{\mu}[\,\omega\!=\!0\,]=\hat{a}^{(0)}_{\mu}+\delta^{WZ}_{\xi}\,\hat{a}^{(0)}_{\mu},\;
\hat{\lambda}'_\a[\,\omega\!=\!0\,]=\hat{\lambda}^{(0)}_\a+\delta^{WZ}_{\xi}\,\hat{\lambda}^{(0)}_\a,\;
\hat{D}'[\,\omega\!=\!0\,]=\hat{D}^{(0)}+\delta^{WZ}_{\xi}\,\hat{D}^{(0)}.
\end{equation*}
Note that $\hat{a}^{(0)}_{\mu}=\hat{a}_{\mu}[\,\omega\!=\!0\,]$,
$\hat{\lambda}^{(0)}_\a=\hat{\lambda}_\a[\,\omega\!=\!0\,]$ and
$\hat{D}^{(0)}=\hat{D}[\,\omega\!=\!0\,]$, and also note that 
$\delta^{WZ}_{\xi}$ gives  --just set  $\omega^{\mu\nu}=0$-- the ordinary supersymmetry transformations in the
Wess-Zumino gauge in~(\ref{susycomp}). The
reader should bear in mind  that
$\hat{\Omega}[a_{\mu},\Omega,\theta]$ is the same for the fields in
$(\hat{a}_{\mu},\hat{\lambda}_{\a}, \hat{D})$ as for their transformed  fields
$\hat{a}'_{\mu},\lambda'_{\a}$ and $D'$ in~(\ref{compprime}). It is thus
clear that imposing invariance under the noncommutative supersymmetry
transformations in~(\ref{susycomp}) will be compatible with demanding
ordinary gauge invariance, and, hence, with asking for noncommutative gauge invariance for
SO(10).

Now, using de definitions in~(\ref{susycomp}), it is not difficult to show that
\begin{equation}
[\hat{\delta}^{WZ}_{\xi_1},\hat{\delta}^{WZ}_{\xi_2}]\hat{{\cal X}}=
-2\,i\,(\xi_1\sigma^{\mu}\bar{\xi}_2-\xi_2\sigma^{\mu}\bar{\xi}_1)\,\partial_{\mu}
\,\hat{{\cal X}}\,+\,\hat{\delta}_{\tilde{\Lambda}}\,\hat{{\cal X}},
\label{susyrep}
\end{equation}
where $\hat{{\cal X}}$ stands for any of the fields in
$(\hat{a}_{\mu},\hat{\lambda}_{\a},D)$, $\tilde{\Lambda}$ is given by 
\begin{equation}
\tilde{\Lambda}=
2\,i\,(\xi_1\sigma^{\mu}\bar{\xi}_2-\xi_2\sigma^{\mu}\bar{\xi}_1)\,\hat{a}_{\mu}
\label{tildelambda}
\end{equation}
and 
\begin{equation*}
\hat{\delta}_{\tilde{\Lambda}}\,\hat{a}_{\mu}=
\partial_{\mu}\,\tilde{\Lambda}+i[\hat{a}_{\mu},\tilde{\Lambda}]_{\star},\;
\hat{\delta}_{\tilde{\Lambda}}\,\hat{\lambda}_{\mu}=i[\hat{\lambda}_{\a},\tilde{\Lambda}]_{\star},\;\hat{\delta}_{\tilde{\Lambda}}\,\hat{D}=[\hat{D},\tilde{\Lambda}]_{\star}
\end{equation*}
are noncommutative gauge transformations. From equation~(\ref{susyrep}) one
draws the conclusion that the space of solutions,
$(\hat{a}_{\mu},\hat{\lambda}_{\a},\hat{D})$, of the Seiberg-Witten map
equations in~(\ref{sweqgauge}) carries a representation of the ${\cal N}=1$
supersymmetry algebra; a representation which is linear modulo noncommutative
gauge transformations.

We are now ready to introduce the noncommutative superYang-Mills action,
$S_{NCSYM}$, of our noncommutative new minimal and minimal supersymmetric
Grand Unified Theories. Firstly, we restrict ourselves to  solutions
$\hat{\Omega}$, $\hat{a}_{\mu}$ $\hat{\lambda}_{\a}$  and $\hat{D}$ to~(\ref{sweqgauge}) which satisfy
\begin{equation*}
\hat{\Omega}[\,\omega\!=\!0\,]=\Omega,\;\hat{a}_{\mu}[\,\omega\!=\!0\,]=a_{\mu},\;
\hat{\lambda}_{\a}[\,\omega\!=\!0\,]=\lambda_{\a},\;\bar{\hat{\lambda}}_{\dalpha}[\,\omega\!=\!0\,]=
\bar{\lambda}_{\dalpha},\;\hat{D}[\,\omega\!=\!0\,]=D.
\end{equation*}
Secondly,  we use this triplet $(\hat{a}_{\mu},\hat{\lambda}_{\a},
\hat{D}) $ and equation~(\ref{wzsvector}) to construct the corresponding
noncommutative $\hat{V}$, with noncommutative field strength given by
\begin{equation*}
\hat{W}_{\a}=-\frac{1}{4}\,\bar{D}^2(e^{-2\hat{V}}D_\alpha e^{2\hat{V}}).
\end{equation*}
Finally, $S_{NCSYM}$ is defined as follows:
\begin{equation}
S_{NCSYM}=\frac{1}{64\pi}{\rm Im}\Big\{\tau\idx\,d^2\theta\,\hat{W}^\alpha
 \hat{W}_{\alpha}\Big\},
\label{ncsymact}
\end{equation}
where $\tau=\frac{\theta_{YM}}{2\pi}+\frac{4\pi i}{g^2}$.
$S_{NCSYM}$ is manifestly invariant under the noncommutative supersymmetry
transformation in~(\ref{ncsusytrans})  and the noncommutative gauge
transformation in~(\ref{supergt}). Obviously, one reaches the same
conclusion is one expresses first $S_{NCSYM}$ in terms of the fields in the
noncommutative supermultiplet $(\hat{a}_{\mu},\hat{\lambda}_{\a},\hat{D})$ and then
one uses~(\ref{susycomp}) and~(\ref{ncbrsop}).

\subsection{The Noncommutative Matter and Noncommuative Higgs Superfields and their Interactions}

In this subsection we shall apply the ideas put forward in the previous
section to the construction  of the noncommutative superfields that we shall
take as the noncommutative counterparts of the ordinary matter superfields
$\Phi^{(16)}_f$, $f=1,2,3$, and the ordinary Higgs superfileds $\Phi^{(H)}$,
$H=210,\,10,\,126,\,\overline{126},\,120$,  in~(\ref{mattercomp}). Then, we
shall easily built their noncommutative interactions with the  vector
superfield of the previous subsection and also construct the noncommutative superpotential. Thus 
we shall generalize $S_{V\Phi}$ and $S_{spot}$ in~(\ref{sactions}) to the noncommutative case.

Let us introduce the following chiral superfields
\begin{equation}
\begin{array}{l}
{\hat{\Phi}^{(16)}_f=\hat{A}^{(16)}_f(y)+\sqrt{2}\theta\hat{\psi}^{(16)}_f(y)+\theta^2
  \hat{F}^{(16)}_f(y),\;f=1,2,3,}\\[4pt]
{\hat{\tilde{\Phi}}^{(16)}_f=\hat{\tilde{A}}^{(16)}_f(y)+\sqrt{2}\theta\hat{\tilde{\psi}}^{(16)}_f(y)+\theta^2
  \hat{\tilde{F}}^{(16)}_f(y),\;f=1,2,3,}\\[4pt]
{\hat{\Phi}^{(H)}=\hat{A}^{(H)}(y)+\sqrt{2}\theta\hat{\psi}^{(H)}(y)+\theta^2 \hat{F}^{(H)}(y),\quad
  H=210,10,126,\overline{126},120,}
\end{array}
\label{ncmattersf}
\end{equation}
where $\hat{A}^{(16)}_f$, $\hat{\psi}^{(16)}_f$, $\hat{F}^{(16)}_f$,
$\hat{\tilde{A}}^{(16)}_f$, $\hat{\tilde{\psi}}^{(16)}_f$,
$\hat{\tilde{F}}^{(16)}_f$,
$\hat{A}^{(H)}$, $\hat{\psi}^{(H)}$ and $\hat{F}^{(H)}$ are
noncommutative fields, which we shall define below  by using the enveloping-algebra formalism
of Refs.~\cite{Madore:2000en, Jurco:2001rq, Aschieri:2002mc}.

Firstly,
$\hat{A}^{(16)}_f,$$\hat{\psi}^{(16)}_f$ and $\hat{F}^{(16)}_f$ are
functions of the corresponding ordinary fields, $A^{(16)}_f,$$\psi^{(16)}_f$
and $F^{(16)}_f(x)$ --see~(\ref{mattercomp})--,
the ordinary gauge field $a_{\mu}$ --see~(\ref{ordsmult})-- and
$\omega^{\mu\nu}$ that solve the following Seiberg-Witten equations in
BRS form:
\begin{equation}
\hat{s}\,\hat{A}^{(16)}_f=s\,\hat{A}^{(16)}_f,\;
\hat{s}\,\hat{\psi}^{(16)}_f=s\,\hat{\psi}^{(16)}_f,\;
\hat{s}\,\hat{F}^{(16)}_f=s\,\hat{F}^{(16)}_f.
\label{sw16}
\end{equation}
The action of the BRS operators  $\hat{s}$ --noncommutative-- and $s$
--ordinary-- on the corresponding fields is defined as follows:
\begin{equation}
\begin{array}{l}
{\hat{s}\,\hat{A}^{(16)}_f=-i\,\hat{\Omega}\star\hat{A}^{(16)}_f,\;
\hat{s}\,\hat{\psi}^{(16)}_f=-i\,\hat{\Omega}\star\hat{\psi}^{(16)}_f,\;
\hat{s}\,\hat{F}^{(16)}_f=-i\,\hat{\Omega}\star\hat{F}^{(16)}_f,}\\[4pt]
{s\,A^{(16)}_f=-i\,\Omega\,A^{(16)}_f,\;
s\,\psi^{(16)}_f=-i\,\Omega\,\psi^{(16)}_f,\;
s\,F^{(16)}_f=-i\,\Omega\,\hat{F}^{(16)}_f,}
\end{array}
\label{hatBRS16}
\end{equation}
where $\hat{\Omega}$ is the very same noncommutative object which occurs
in~(\ref{ncbrsop}).

Secondly, $\hat{\tilde{A}}^{(16)}_f,$$\hat{\tilde{\psi}}^{(16)}_f$ and $\hat{\tilde{F}}^{(16)}_f$ are also
functions of the corresponding ordinary fields, $\tilde{A}^{(16)}_f,$$\tilde{\psi}^{(16)}_f$
and $\tilde{F}^{(16)}_f(x)$ --see~(\ref{mattercomp})--,
the ordinary gauge field $a_{\mu}$ --see~(\ref{ordsmult})-- and
$\omega^{\mu\nu}$ which satisfy
\begin{equation}
\hat{s}\,\hat{\tilde{A}}^{(16)}_f=s\,\hat{\tilde{A}}^{(16)}_f,\;
\hat{s}\,\hat{\tilde{\psi}}^{(16)}_f=s\,\hat{\tilde{\psi}}^{(16)}_f,\;
\hat{s}\,\hat{\tilde{F}}^{(16)}_f=s\,\hat{\tilde{F}}^{(16)}_f.
\label{sw16b}
\end{equation}
The BRS operators $\hat{s}$  and $s$ act thus on the corresponding fields in
the previous set of equation:
\begin{equation}
\begin{array}{l}
{\hat{s}\,\hat{\tilde{A}}^{(16)}_f=i\,\hat{\tilde{A}}^{(16)}_f\star\hat{\Omega},\;
\hat{s}\,\hat{\tilde{\psi}}^{(16)}_f=i\,\hat{\tilde{\psi}}^{(16)}_f\star\hat{\Omega},\;
\hat{s}\,\hat{\tilde{F}}^{(16)}_f=i\,\hat{\Omega}\star\hat{\tilde{F}}^{(16)}_f\star\hat{\Omega},}\\[4pt]
{s\,\tilde{A}^{(16)}_f=i\,\tilde{A}^{(16)}_f\,\Omega,\;
s\,\tilde{\psi}^{(16)}_f=i\,\tilde{\psi}^{(16)}_f\,\Omega,\;
s\,\tilde{F}^{(16)}_f=i\,\tilde{F}^{(16)}_f\,\Omega,}
\end{array}
\label{hatBRS16B}
\end{equation}
where, again, $\hat{\Omega}$ is the very same noncommutative object which enters~(\ref{ncbrsop}).

Finally, $\hat{A}^{(H)},$$\hat{\psi}^{(H)}$ and $\hat{F}^{(H)}$
are functions of the corresponding ordinary fields, $A^{(H)},$$\psi^{(H)}$ and $\hat{F}^{(H)}$, in~(\ref{mattercomp}),
the ordinary gauge field $a_{\mu}$ --see~(\ref{ordsmult})-- and $\omega^{\mu\nu}$  that solve the following Seiberg-Witten map equations in BRS form:
\begin{equation}
\hat{s}\,\hat{A}^{(H)}=s\,\hat{A}^{(H)},\;
\hat{s}\,\hat{\psi}^{(H)}=s\,\hat{\psi}^{(H)},\;
\hat{s}\,\hat{F}^{(H)}=s\,\hat{F}^{(H)},
\label{swH}
\end{equation}
where now
\begin{equation}
\begin{array}{l}
{\hat{s}\,\hat{A}^{(H)}=-i\,[\hat{\Omega},\hat{A}^{(16)}]_{\star},\;
\hat{s}\,\hat{\psi}^{(H)}=-i\,[\hat{\Omega},\hat{\psi}^{(H)}]_{\star},\;
\hat{s}\,\hat{F}^{(H)}=-i\,[\hat{\Omega},\hat{F}^{(H)}]_{\star},}\\[4pt]
{s\,A^{(H)}=-i\,[\Omega,A^{(16)}],\;
s\,\psi^{(H)}=-i\,[\Omega,\psi^{(H)}],\;
s\,F^{(H)}=-i\,[\Omega,F^{(H)}].}
\end{array}
\label{hatBRSH}
\end{equation}

It is plain that the construction of $\hat{\Phi}^{(16)}_f,$
$\hat{\tilde{\Phi}}^{(16)}_f$ and $\hat{\Phi}^{(H)}$ yields noncommutative
superfields that are well-defined on the infinitesimal gauge orbit of
the ordinary fields they are functions of. Indeed, one readily sees that
\begin{equation}
\begin{array}{l}
{\hat{\Phi}^{(16)}_f[a_\mu+\delta_{\Omega}\,a_{\mu};
A^{(16)}_f+\delta_{\Omega}\,A^{(16)}_f;
\psi^{(16)}_f+\delta_{\Omega}\,\psi^{(16)}_f;
F^{(16)}_f+\delta_{\Omega}\,F^{(16)}_f]=
\hat{\Phi}^{(16)}+\hat{\delta}_{\hat{\Omega}}\,\hat{\Phi}^{(16)},}\\[4pt]
{\hat{\tilde{\Phi}}^{(16)}_f[a_\mu+\delta_{\Omega}\,a_{\mu};
\tilde{A}^{(16)}_f+\delta_{\Omega}\,\tilde{A}^{(16)}_f;
\tilde{\psi}^{(16)}_f+\delta_{\Omega}\,\tilde{\psi}^{(16)}_f;
\tilde{F}^{(16)}_f+\delta_{\Omega}\,\tilde{F}^{(16)}_f]=
\hat{\tilde{\Phi}}^{(16)}+\hat{\delta}_{\hat{\Omega}}\,\hat{\tilde{\Phi}}^{(16)},}\\[4pt]
{\hat{\Phi}^{(H)}[a_\mu+\delta_{\Omega}\,a_{\mu};
A^{(H)}+\delta_{\Omega}\,A^{(H)};
\psi^{(H)}+\delta_{\Omega}\,\psi^{(H)};
F^{(H)}+\delta_{\Omega}\,F^{(H)}]=
\hat{\Phi}^{(H)}+\hat{\delta}_{\hat{\Omega}}\,\hat{\Phi}^{(H)},}
\end{array}
\label{matgaugetra}
\end{equation}
where
\begin{equation*}
\hat{\delta}_{\hat{\Omega}}\,\hat{\Phi}^{(16)}=-i\,\hat{\Omega}\star\hat{\Phi}^{(16)},\;
\hat{\delta}_{\hat{\Omega}}\,\hat{\tilde{\Phi}}^{(16)}=i\,\hat{\tilde{\Phi}}^{(16)}\star\hat{\Omega},\;
\hat{\delta}_{\hat{\Omega}}\,\hat{\Phi}^{(H)}=-i\,[\hat{\Omega},\hat{\Phi}^{(H)}]_{\star}.
\end{equation*}

In~(\ref{ncsusytrans}), we have defined the action of the noncommutative supersymmetry operator in the
Wess-Zumino gauge, $\hat{\delta}^{WZ}_{\xi}$, on the vector superfield $\hat{V}$ . This definition leads to the
following action of
$\hat{\delta}^{WZ}_{\xi}$ on $\hat{\Phi}^{(16)}_f$,
$\hat{\tilde{\Phi}}^{(16)}_f$ and $\hat{\Phi}^{(H)}$:
\begin{equation}
\begin{array}{l}
{\hat{\delta}^{WZ}_{\xi}\,\hat{\Phi}^{(16)}_f=(\xi
Q+\bar{\xi}\bar{Q})\,\hat{\Phi}^{(16)}_f+\hat{\delta}_{\hat{\Lambda}}\,\hat{\Phi}^{(16)}_f,\quad
\hat{\delta}_{\hat{\Lambda}}\,\hat{\Phi}^{(16)}_f=-i\,\hat{\Lambda}\star\hat{\Phi}^{(16)}_f,}\\[4pt]
{\hat{\delta}^{WZ}_{\xi}\,\hat{\tilde{\Phi}}^{(16)}_f=(\xi
Q+\bar{\xi}\bar{Q})\,\hat{\tilde{\Phi}}^{(16)}_f+\hat{\delta}_{\hat{\Lambda}}\,\hat{\tilde{\Phi}}^{(16)}_f,\quad
\hat{\delta}_{\hat{\Lambda}}\,\hat{\tilde{\Phi}}^{(16)}_f=i\,\hat{\tilde{\Phi}}^{(16)}_f}\star\hat{\Lambda,}\\[4pt]
{\hat{\delta}^{WZ}_{\xi}\,\hat{\Phi}^{(H)}=(\xi
Q+\bar{\xi}\bar{Q})\,\hat{\Phi}^{(H)}+\hat{\delta}_{\hat{\Lambda}}\,\hat{\Phi}^{(H)},\quad
\hat{\delta}_{\hat{\Lambda}}\,\hat{\Phi}^{(H)}=-i\,[\hat{\Lambda},\hat{\Phi}^{(H)}]_{\star}},
\end{array}
\label{mattsfieltra}
\end{equation}
where $\hat{\Lambda}$ is the chiral noncommutative superfield given in~(\ref{hatlambda}).

The action of $\hat{\delta}^{WZ}_{\xi}$ on the components of the noncommutative
matter and Higgs superfields can be worked out from~(\ref{mattsfieltra}). One obtains thus 
\begin{equation}
\begin{array}{l}
{\hat{\delta}^{WZ}_{\xi}\,\hat{A}^{(16)}_f=\sqrt{2}\,\xi\hat{\psi}^{(16)}_f,\quad
\hat{\delta}^{WZ}_{\xi}\,\hat{\psi}^{(16)}_{\a\,f}=\sqrt{2}\,i\,(\sigma^{\mu}\bar{\xi})_\a\,{\cal
  D}_{\mu}\,\hat{A}^{(16)}_f+\xi_\a\,\hat{F}^{(16)}_f,}\\[4pt]
{\hat{\delta}^{WZ}_{\xi}\,\hat{F}^{(16)}_f=i\,\sqrt{2}\,\bar{\xi}\bar{\sigma}^{\mu}{\cal
    D}_{\mu}\hat{\psi}^{(16)}_f+ 2\,i\,\bar{\xi}\bar{\hat{\lambda}}\star\hat{A}^{(16)}_f,}\\[6pt]
{\hat{\delta}^{WZ}_{\xi}\,\hat{\tilde{A}}^{(16)}_f=\sqrt{2}\,\xi\hat{\tilde{\psi}}^{(16)}_f,\quad
\hat{\delta}^{WZ}_{\xi}\,\hat{\tilde{\psi}}^{(16)}_{\a\,f}=\sqrt{2}\,i\,(\sigma^{\mu}\bar{\xi})_\a\,{\cal
  D}_{\mu}\,\hat{\tilde{A}}^{(16)}_f+\xi_\a\,\hat{\tilde{F}}^{(16)}_f,}\\[4pt]
{\hat{\delta}^{WZ}_{\xi}\,\hat{\tilde{F}}^{(16)}_f=i\,\sqrt{2}\,\bar{\xi}\bar{\sigma}^{\mu}{\cal
    D}_{\mu}\hat{\tilde{\psi}}^{(16)}_f-
  2\,i\,\hat{\tilde{A}}^{(16)}_f\star\bar{\xi}\bar{\hat{\lambda}},}\\[6pt]
{\hat{\delta}^{WZ}_{\xi}\,\hat{A}^{(H)}=\sqrt{2}\,\xi\hat{\psi}^{(H)},\quad
\hat{\delta}^{WZ}_{\xi}\,\hat{\psi}^{(H)}_{\a}=\sqrt{2}\,i\,(\sigma^{\mu}\bar{\xi})_\a\,{\cal
  D}_{\mu}\,\hat{A}^{(H)}+\xi_\a\,\hat{F}^{(H)},}\\[4pt]
{\hat{\delta}^{WZ}_{\xi}\,\hat{F}^{(H)}=i\,\sqrt{2}\,\bar{\xi}\bar{\sigma}^{\mu}{\cal
    D}_{\mu}\hat{\psi}^{(H)} + 2\,i\,[\bar{\xi}\bar{\hat{\lambda}},\hat{A}^{(H)}]_{\star},}
\end{array}
\label{mattercomptrans}
\end{equation}
where
\begin{equation*}
\begin{array}{l}
{{\cal D}_{\mu}\hat{A}^{(16)}_f=\partial_{\mu}\hat{A}^{(16)}_f+\,i\,\hat{a}_{\mu}\star\hat{A}^{(16)}_f,\,
{\cal D}_{\mu}\hat{\psi}^{(16)}_f=\partial_{\mu}\hat{\psi}^{(16)}_f+\,i\,\hat{a}_{\mu}\star\hat{\psi}^{(16)}_f,}\\[4pt]
{{\cal D}_{\mu}\hat{\tilde{A}}^{(16)}_f=\partial_{\mu}\hat{\tilde{A}}^{(16)}_f-\,i\hat{\tilde{A}}^{(16)}_f\star\hat{a}_{\mu},\,
{\cal D}_{\mu}\hat{\tilde{\psi}}^{(16)}_f=\partial_{\mu}\hat{\tilde{\psi}}^{(16)}_f-\,i\,\hat{\tilde{\psi}}^{(16)}_f\star\hat{a}_{\mu},}\\[4pt]
{{\cal D}_{\mu}\hat{A}^{(H)}=\partial_{\mu}\hat{A}^{(H)}+\,i\,[\hat{a}_{\mu},\hat{A}^{(H)}]_{\star},\,
{\cal
  D}_{\mu}\hat{\psi}^{(H)}=\partial_{\mu}\hat{\psi}^{(H)}_f+\,i\,[\hat{a}_{\mu},\hat{\psi}^{(H)}]_{\star}.}
\end{array}
\end{equation*}
Let us recall --see~(\ref{susycomp}) and~(\ref{deltaclass})-- that the
action of $\hat{\delta}^{WZ}_{\xi}$ on the components of $\hat{V}$ is  well-defined
on the infinitesimal gauge orbit of the ordinary fields these components
depend upon. This state of affairs also occurs for the components of the
noncommutative matter, $\hat{\Phi}^{(16)}_f$ and
$\hat{\tilde{\Phi}}^{(16)}_f$, and Higgs superfields, $\hat{\Phi}^{(H)}$, $H=210, 10, 126,
\overline{126}, 120$, constructed above. Indeed, if $\hat{\varphi}$ stands for any of those noncommutative components, 
then
\begin{equation*}
\hat{\delta}^{WZ}_{\xi}\,\hat{\varphi}[a_{\mu}+\delta_{\Omega}\,a_{\mu};\varphi+\delta_{\Omega}\,\varphi]=
\hat{\delta}_{\hat{\Omega}}\,(\hat{\delta}^{WZ}_{\xi}\,\hat{\varphi}),
\end{equation*}
where
$\hat{\delta}_{\hat{\Omega}}\,(\hat{\delta}^{WZ}_{\xi}\,\hat{\varphi})=-i\,\hat{\Omega}\star\hat{\delta}^{WZ}_{\xi}\,\hat{\varphi}$,
for the components of $\hat{\Phi}^{(16)}_f$;
$\hat{\delta}_{\hat{\Omega}}\,(\hat{\delta}^{WZ}_{\xi}\,\hat{\varphi})=-i\,\hat{\delta}^{WZ}_{\xi}\,\hat{\varphi}\star\hat{\Omega}$,
if $\hat{\varphi}$ denotes any component of $\hat{\tilde{\Phi}}^{(16)}_f$;
and
$\hat{\delta}_{\hat{\Omega}}\,(\hat{\delta}^{WZ}_{\xi}\,\hat{\varphi})=-i\,[\hat{\Omega},\hat{\delta}^{WZ}_{\xi}\,\hat{\varphi}]_{\star}$,
  when they are  the  components of $\hat{\Phi}^{(H)}$ the ones we are dealing
  with. $\delta_{\Omega}$ generates the ordinary infinitesimal gauge
  transformations.
We then conclude that there is no obstruction  to demand  gauge invariance and
  invariance under the supersymmetry transformations in~(\ref{mattercomptrans}) at the same time.

Let $\hat{\varphi}$ denote again any of the noncommutative component fields in
$(\hat{A}^{(16)}_f,\hat{\psi}^{(16)}_f,\hat{F}^{(16)}_f)$,
$(\hat{\tilde{A}}^{(16)}_f,\hat{\tilde{\psi}}^{(16)}_f,\hat{\tilde{F}}^{(16)}_f)$
or $(\hat{A}^{(H)},\hat{\psi}^{(H)},\hat{F}^{(H)})$. Then, using the fact that
$\hat{\varphi}$ solves the appropriate Seiberg-Witten map equations in~(\ref{sw16}),~(\ref{sw16b}) or~(\ref{swH}), it is not difficult to show
that
\begin{equation*}
\hat{\varphi}'=\hat{\varphi}+\hat{\delta}^{WZ}_{\xi}\,\hat{\varphi}
\end{equation*}
solves the same Seiberg-Witten map equation as $\hat{\varphi}$. Of
course,  at $\omega^{\mu\nu}=0$, $\hat{\varphi}'$  and $\hat{\varphi}$ differ by
an ordinary supersymmetry transformation in the Wess-Zumino gauge of
$\hat{\varphi}[\,\omega\!=0\!\,]$.

We have seen that the spaces of solutions of the Seiberg-Witten map equations
in~(\ref{sw16}),~(\ref{sw16b}) and~(\ref{swH}) are constituted,
respectively, by the noncommutative matter,
$(\hat{A}^{(16)}_f,\hat{\psi}^{(16)}_f,\hat{F}^{(16)}_f)$,
$(\hat{\tilde{A}}^{(16)}_f,\hat{\tilde{\psi}}^{(16)}_f,\hat{\tilde{F}}^{(16)}_f)$,
and Higgs, $(\hat{A}^{(H)},\hat{\psi}^{(H)},\hat{F}^{(H)})$, triplets. On these spaces of solutions
$\hat{\delta}^{WZ}_{\xi}$ acts according to the formulae in~(\ref{mattercomptrans}).
Let us show now that each of these  spaces of solutions carries a representation
of the ${\cal N}=1$ supersymmetry algebra.  Taking into account the
definitions in~(\ref{mattercomptrans}), one shows that
\begin{equation*}
[\hat{\delta}^{WZ}_{\xi_1},\hat{\delta}^{WZ}_{\xi_2}]\hat{\varphi}=
-2\,i\,(\xi_1\sigma^{\mu}\bar{\xi}_2-\xi_2\sigma^{\mu}\bar{\xi}_1)\,\partial_{\mu}
\,\hat{\varphi}\,+\,\hat{\delta}_{\tilde{\Lambda}}\,\hat{\varphi},
\end{equation*}
where $\hat{\varphi}$ stands for any of the fields in the noncommutative
matter and Higgs triplets we are dealing with and
\begin{equation*}
\tilde{\Lambda}=
2\,i\,(\xi_1\sigma^{\mu}\bar{\xi}_2-\xi_2\sigma^{\mu}\bar{\xi}_1)\,\hat{a}_{\mu}.
\end{equation*}
Of course, this is the same $\tilde{\Lambda}$ as for the noncommutative gauge
supermultiplet $(\hat{a}_{\mu},\hat{\lambda}_\a,\hat{D})$:
see equations~(\ref{susyrep}) and~(\ref{tildelambda}).
The noncommutative gauge transformation
$\hat{\delta}_{\tilde{\Lambda}}\,\hat{\varphi}$ is given by
\begin{equation*}
\begin{array}{l}
{\hat{\delta}_{\tilde{\Lambda}}\,\hat{\varphi}=-i\,\tilde{\Lambda}\star\hat{\varphi},\quad\text{if}\quad
\hat{\varphi}\in(\hat{A}^{(16)}_f,\hat{\psi}^{(16)}_f,\hat{F}^{(16)}_f),}\\[4pt]
{\hat{\delta}_{\tilde{\Lambda}}\,\hat{\varphi}=i\hat{\varphi}\star\tilde{\Lambda},\quad\text{if}\quad
\hat{\varphi}\in(\hat{\tilde{A}}^{(16)}_f,\hat{\tilde{\psi}}^{(16)}_f,\hat{\tilde{F}}^{(16)}_f),}\\[4pt]
{\hat{\delta}_{\tilde{\Lambda}}\,\hat{\varphi}=-i[\tilde{\Lambda},\hat{\varphi}]_{\star},
\quad\text{if}\quad
\hat{\varphi}\in(\hat{A}^{(H)},\hat{\psi}^{(H)},\hat{F}^{(H)})}.
\end{array}
\end{equation*}

We are now ready to introduce the noncommutative deformations, say
$S_{\hat{V}\hat{\Phi}}$ and $S_{\widehat{spot}}$, of
$S_{V\Phi}$ and $S_{spot}$ in~(\ref{sactions}). But first, we impose the
following conditions on the components of the noncommutative matter superfields
$\hat{\Phi}^{(16)}_f$, $\hat{\tilde{\Phi}}^{(16)}_f$ and $\hat{\Phi}^{(H)}$:
\begin{equation}
\begin{array}{l}
{\hat{A}^{(16)}_f[\,\omega\!=\!0\,]=A^{(16)}_f,\quad
\hat{\psi}^{(16)}_f[\,\omega\!=\!0\,]=\psi^{(16)}_f,\quad
\hat{F}^{(16)}_f[\,\omega\!=\!0\,]=F^{(16)}_f,}\\[4pt]
{\hat{\tilde{A}}^{(16)}_f[\,\omega\!=\!0\,]=\tilde{A}^{(16)}_f,\quad
\hat{\tilde{\psi}}^{(16)}_f[\,\omega\!=\!0\,]=\tilde{\psi}^{(16)}_f,\quad
\hat{\tilde{F}}^{(16)}_f[\,\omega\!=\!0\,]=\tilde{F}^{(16)}_f,}\\[4pt]
{\hat{A}^{(H)}[\,\omega\!=\!0\,]=A^{(H)},\quad
\hat{\psi}^{(H)}_f[\,\omega\!=\!0\,]=\psi^{(H)},\quad
\hat{F}^{(H)}[\,\omega\!=\!0\,]=F^{(H)}.}
\end{array}
\label{matcond}
\end{equation}
Furnished with the noncommutative superfields
$\hat{\Phi}^{(16)}_f,$$\hat{\tilde{\Phi}}^{(16)}_f$ and $\hat{\Phi}^{(H)}$
whose components satisfy the conditions in~(\ref{matcond}) and the
noncommutative vector superfield $\hat{V}$ employed to define $S_{NCSYM}$ in~(\ref{ncsymact}), we  define  $S_{\hat{V}\hat{\Phi}}$ and
$S_{\widehat{spot}}$ in terms of them as follows
\begin{equation}
\begin{array}{l}
{S_{\hat{V}\hat{\Phi}}\,=\,\idx\,d^2\theta\,d^2\bar{\theta}\;
 \sum_{f}\,(\hat{\Phi}^{(16)})^{\dagger}_f\star e^{2\hat{V}}\star\hat{\Phi}^{(16)}_f\,+\,\sum_{H}\,\frac{1}{s(H)}{\rm
   Tr}\Big((\hat{\Phi}^{(H)})^{\dagger}\star
 e^{2\hat{V}}\star\hat{\Phi}^{(H)}\star e^{-2\hat{V}}\Big),}\\[4pt]
{S_{\widehat{spot}}=\idx\,d^2\theta\;\{ W_{\widehat{matter}}\,+\,W_{\widehat{Higgs}}\}\,+\,{\rm h.c.},}
\end{array}
\label{ncsactions}
\end{equation}
where
\begin{equation}
\begin{array}{l}
{ W_{\widehat{matter}}=\sum_{f,f'}\;\Big\{{\cal
      Y}^{(10)}_{ff'}\;\hat{\tilde{\Phi}}^{(16)}_f\star\hat{\Phi}^{(10)}\star\hat{\Phi}^{(16)}_{f'}}\\[4pt]
{\phantom{W_{\widehat{matter}}=}+{\cal
      Y}^{(\overline{126})}_{ff'}\;\hat{\tilde{\Phi}}^{(16)}_f\star\hat{\Phi}^{(\overline{126})}\star{\Phi}^{(16)}_{f'}+

{\cal
      Y}^{(120)}_{ff'}\;\hat{\tilde{\Phi}}^{(16)}_f\star\hat{\Phi}^{(120)}\star\hat{\Phi}^{(16)}_{f'}\Big\}}\\[4pt]
{\text{and}}\\[4pt]
{W_{\widehat{Higgs}}= \frac{M^{(210)}}{64 (4!)^2}{\rm Tr}\,\hat{\Phi}^{(210)}\star\hat{\Phi}^{(210)}
           -\frac{M^{(126)}}{32 (5!)^2}{\rm Tr}\,\hat{\Phi}^{(\overline{126})}\star\hat{\Phi}^{(126)}
           + \frac{M^{(10)}}{64 }{\rm
             Tr}\,\hat{\Phi}^{(10)}\star\hat{\Phi}^{(10)}}\\[4pt]
{\phantom{W_{\widehat{Higgs}}=}
           -\frac{M^{(120)}}{64 (3!)^2}{\rm
             Tr}\,\hat{\Phi}^{(120)}\star\hat{\Phi}^{(120)}+\lambda_1\,{\rm
             Tr}\,\hat{\Phi}^{(210)}\star\hat{\Phi}^{(210)}\star\hat{\Phi}^{(210)}}\\[4pt]
{\phantom{W_{\widehat{Higgs}}=}
+\lambda^{(1)}_2 {\rm Tr}\,\hat{\Phi}^{(210)}\star\hat{\Phi}^{(\overline{126})}\star\hat{\Phi}^{(126)}+\lambda^{(2)}_2 {\rm Tr}\,\hat{\Phi}^{(210)}\star\hat{\Phi}^{(126)}\star\hat{\Phi}^{(\overline{126})}}\\[4pt]
{\phantom{W_{\widehat{Higgs}}=}+\lambda^{(1)}_3 {\rm Tr}\,\hat{\Phi}^{(10)}\star\hat{\Phi}^{(120)}\star\hat{\Phi}^{(210)}+\lambda^{(2)}_3 {\rm Tr}\,\hat{\Phi}^{(10)}\star\hat{\Phi}^{(210)}\star\hat{\Phi}^{(120)}}\\[4pt]
{\phantom{W_{Higgs}=}+\lambda^{(1)}_4\,{\rm Tr}\,\hat{\Phi}^{(120)}\star\hat{\Phi}^{(210)}\star\hat{\Phi}^{(126)}+ \lambda^{(2)}_4\,{\rm Tr}\,\hat{\Phi}^{(120)}\star\hat{\Phi}^{(126)}\star\hat{\Phi}^{(210)}}\\[4pt]
{\phantom{W_{Higgs}=}+\lambda^{(1)}_5 {\rm Tr}\,\hat{\Phi}^{(10)}\star\hat{\Phi}^{(210)}\star\hat{\Phi}^{(126)}+\lambda^{(2)}_5 {\rm Tr}\,\hat{\Phi}^{(10)}\star\hat{\Phi}^{(126)}\star\hat{\Phi}^{(210)}}\\[4pt]
{\phantom{W_{Higgs}=}+\lambda^{(1)}_6 {\rm Tr}\,\hat{\Phi}^{(10)}\star\hat{\Phi}^{(210)}\star\hat{\Phi}^{(\overline{126})}+\lambda^{(2)}_6 {\rm Tr}\,\hat{\Phi}^{(10)}\star\hat{\Phi}^{(\overline{126})}\star\hat{\Phi}^{(210)}}\\[4pt]
{\phantom{W_{Higgs}=}+\lambda_7\,{\rm
      Tr}\,\hat{\Phi}^{(120)}\star\hat{\Phi}^{(120)}\star\hat{\Phi}^{(210)}}\\[4pt]
{\phantom{W_{Higgs}=}+\lambda^{(1)}_8 {\rm
    Tr}\,\hat{\Phi}^{(120)}\star\hat{\Phi}^{(210)}\star\hat{\Phi}^{(\overline{126})}+\lambda^{(2)}_8 {\rm
    Tr}\,\hat{\Phi}^{(120)}\star\hat{\Phi}^{(\overline{126})}\star\hat{\Phi}^{(210)}.}
\end{array}
\label{ncww}
\end{equation}

It is apparent that $S_{\hat{V}\hat{\Phi}}$ and $S_{\widehat{spot}}$ are
invariant under the noncommutative supersymmetry transformations
in~(\ref{mattsfieltra}) and the gauge transformations in~(\ref{matgaugetra}). When $S_{\hat{V}\hat{\Phi}}$ and $S_{\widehat{spot}}$
are expressed in terms of the components of the noncommutative superfields, the corresponding
invariance is given by the transformations in~(\ref{susycomp}) and~(\ref{mattercomptrans}), on the one hand, and~(\ref{ncbrsop}),~(\ref{hatBRS16}),~(\ref{hatBRS16B}) and~(\ref{hatBRSH}), on the other hand.

We  would like to stress that $S_{\hat{V}\hat{\Phi}}$ and
$S_{\widehat{spot}}$ in~(\ref{ncsactions}) and~(\ref{ncww}) almost look like the naive deformations of their corresponding
ordinary counterparts $S_{V\Phi}$ and $S_{spot}$, which are displayed in~(\ref{sactions}) and~(\ref{ordspoten}). This likeness we have pointed out  partially stems from
the fact that the components of the Higgs superfields, $\hat{\Phi}^{(H)}$, in~(\ref{ncmattersf}) take values in the Clifford
algebra  $\mathbb{C}\rm{l}_{10}(\mathbb{C})$.
Notice that the doubling that occurs in some of the terms in
$W_{\widehat{Higgs}}$ is due to the fact that given three functions
$f_1$, $f_2$ and $f_3$, then
\begin{equation*}
\idx\;f_1\star f_2\star f_3 \neq \idx\;f_1\star f_3\star f_2,
\end{equation*}
unless two of them are equal.

Finally,  it is easy   --although lengthy-- to express $S_{\hat{V}\hat{\Phi}}$ and $S_{\widehat{spot}}$ in terms
of the ordinary fields. To do so, one first  obtains explicit expressions for
$\hat{A}^{(16)}_f,$$\hat{\psi}^{(16)}_f,$$\hat{F}^{(16)}_f,$$\hat{\tilde{A}}^{(16)}_f,$$\hat{\tilde{\psi}}^{(16)}_f,$$\hat{\tilde{F}}^{(16)}_f,$
$\hat{A}^{(H)},$$\hat{\psi}^{(H)}$ and $\hat{F}^{(H)}$ in terms of
the corresponding ordinary fields: the reader has only to particularize the general expressions in Ref.~\cite{Martin:2012aw} to the
case at hand. Then, one  
substitutes those expressions in~(\ref{ncsactions}) and~(\ref{ncww}) and does 
the lengthy arithmetic.

\section{Final Comments}

In this paper we have formulated the minimal and new minimal supersymmetric
GUTs on canonical (i.e., $[X^{\mu},X^{\nu}]=i\omega^{\mu\nu}$) noncommutative
space-time by using the enveloping-algebra formalism. Taking advantage of the
Seiberg-Witten map,  we have constructed noncommutative superfields in the
Wess-Zumino gauge out of the  ordinary components of the corresponding
ordinary superfields. Thus supersymmetry is linearly realized explictly in
terms of the noncommutative fields. However,  unlike in the $U(n)$ case in
the fundamental representation, the noncommutative supersymmetry
transformations in~(\ref{susycomp}) cannot be
generated by applying the Seiberg-Witten map to an $\omega$-deformed
transformation of the ordinary fields. Indeed, it can be shown --as in  Ref.~\cite{Martin:2008xa}-- that
the equations
\begin{equation*}
\begin{array}{l}
{\hat{a}_{\mu}+\hat{\delta}^{WZ}_{\xi}\,\hat{a}_{\mu}=
\hat{a}_{\mu}[a_{\mu}+\hat{\delta}_{\xi}\,a_{\mu},
\lambda_{\a}+\hat{\delta}_{\xi}\,\lambda_{\a},D+\hat{\delta}_{\xi}\,D],}\\[4pt]
{\hat{\lambda}_{\a}+\hat{\delta}^{WZ}_{\xi}\,\hat{\lambda}_{\a}=
\hat{\lambda}_{\a}[a_{\mu}+\hat{\delta}_{\xi}\,a_{\mu},
\lambda_{\a}+\hat{\delta}_{\xi}\,\lambda_{\a},D+\hat{\delta}_{\xi}\,D],}\\[4pt]
{\hat{D}+\hat{\delta}^{WZ}_{\xi}\,\hat{D}=
\hat{D}[a_{\mu}+\hat{\delta}_{\xi}\,a_{\mu},
\lambda_{\a}+\hat{\delta}_{\xi}\,\lambda_{\a},D+\hat{\delta}_{\xi}\,D]}
\end{array}
\end{equation*}
are not satisfied by any $\hat{\delta}_{\xi}\,\hat{a}_{\mu}$,
$\hat{\delta}_{\xi}\,\lambda_{\a}$ and $\hat{\delta}_{\xi}\,D$ in the Lie
algebra of $SO(10)$, if $\hat{a}_{\mu}[\,\cdot\,,\,\cdot\,,\,\cdot\,],
\hat{\lambda}_{\a}[\,\cdot\,,\,\cdot\,,\,\cdot\,]$ and $\hat{D}[\,\cdot\,,\,\cdot\,,\,\cdot\,]$ define
Seiberg-Witten maps. Analogously, it is not difficult to see that the noncommutative supersymmetry transformations
of the noncommutative Higgsses and their superpartners in~(\ref{mattercomptrans}) cannot be generated from variations of the
corresponding ordinary  fields as follows:
\begin{equation*}
\begin{array}{l}
{\hat{A}^{(H)}+\hat{\delta}^{WZ}_{\xi}\,\hat{A}^{(H)}=
\hat{A}^{(H)}[a_{\mu}+\hat{\delta}_{\xi}\,a_{\mu},
A^{(H)}+\hat{\delta}_{\xi}\,A^{(H)},\psi^{(H)}_\a+\hat{\delta}_{\xi}\,\psi^{(H)}_\a,F^{(H)}+\hat{\delta}_{\xi}\,F^{(H)}],}\\[4pt]
{\hat{\psi}^{(H)}_\a+\hat{\delta}^{WZ}_{\xi}\,\hat{\psi}^{(H)}_\a=
\hat{\psi}^{(H)}_\a[a_{\mu}+\hat{\delta}_{\xi}\,a_{\mu},
A^{(H)}+\hat{\delta}_{\xi}\,A^{(H)},\psi^{(H)}_\a+\hat{\delta}_{\xi}\,\psi^{(H)}_\a,F^{(H)}+\hat{\delta}_{\xi}\,F^{(H)}],}\\[4pt]
{\hat{F}^{(H)}+\hat{\delta}^{WZ}_{\xi}\,\hat{F}^{(H)}=
\hat{F}^{(H)}[a_{\mu}+\hat{\delta}_{\xi}\,a_{\mu},
A^{(H)}+\hat{\delta}_{\xi}\,A^{(H)},\psi^{(H)}_\a+\hat{\delta}_{\xi}\,\psi^{(H)}_\a,F^{(H)}+\hat{\delta}_{\xi}\,F^{(H)}],}
\end{array}
\end{equation*}
where $\hat{A}^{(H)}[\,\cdot\,,\,\cdot\,,\,\cdot\,,\,\cdot\,],
\hat{\psi}^{(H)}_\a[\,\cdot\,,\,\cdot\,,\,\cdot\,,\,\cdot\,]$ and
$\hat{F}^{(H)}[\,\cdot\,,\,\cdot\,,\,\cdot\,,\,\cdot\,]$ give Seiberg-Witten maps.
In summary,  the supersymmetry  of  our noncommutative
SO(10) theories is not realized by the   ordinary
fields, but, recall, it is linearly realized by the noncommutative fields. Let
us point out that in the U(n) case --in the fundamental representation or its siblings--
such realization of the supersymmetry transformations in terms of ordinary
fields exists, but it is at the cost of being a nonlinear $\omega$-dependent
transformation --see~\cite{Martin:2008xa}.

It is thus clear that if one uses ordinary fields --the fields that create and
destroy leptons, quarks, photons, gluons, etc..-- to formulate,
via the Seiberg-Witten map, our SO(10) supersymmetric theories on noncommutative
space-time, the picture that emerges as regards to the  the supersymmetry
properties of those ordinary fields differs radically from the picture that materializes when
those very fields are used to formulate the corresponding supersymmetric
theories on ordinary Minkowski space-time. Indeed, when space-time is noncommutative
there is no supersymmetry in terms of the ordinary fields, although there is a
hidden supersymmetry that reveals itself when the noncommutative fields are
used. It is to early to say whether this absence supersymmetry for the
ordinary fields in the noncommutative theory can be accepted\footnote{I thank
  P. Schupp for raising this issue} as a supersymmetry breaking mechanism relevant for the 
description of Nature: If so, it would be the noncommutative character of space-time
that breaks through interactions the supersymmetry carried by ordinary fields when $\omega^{\mu\nu}=0$. It is cleat that  more understanding of the properties of the theories at
hand is needed before a verdict is issued. It should be noticed that the logarithmic UV/IR mixing phenomena of noncommutative supersymmetric theories~\cite{Ruiz:2000hu}  may be key to
interpreting as a phenonologically relevant supersymmetry breaking mechanism the fact that
supersymmetry is not realized by the ordinary fields in the noncommutative
theory, for otherwise the lower the energy the closer we would be to
$\omega^{\mu\nu}=0$, where supersymmetry is realized (linearly) by the ordinary
fields. Hence, it would seem right to think that in defining the GUTs introduced above the Seiberg-Witten map should not be understood as a formal
power series expansion in $\omega^{\mu\nu}$, but in an $\omega^{\mu\nu}\!$-exact form way--see
Ref.~\cite{Martin:2012aw} for the appropriate formulae. Let us point out that the $\omega\!$-exact
Seiberg-Witten map is not a polynomial in the $\star$-product, so there may be  
UV/IR mixing even though the gauge group is simple.

It is plain that there are many issues --UV/IR mixing, renormalizability, vacua,...-- regarding the noncommutative GUTs we have introduced above that should be studied to gain more understanding of the properties of these theories. In particular,
it is an open problem to see whether our noncommutative GUTs fit in  F-theory --or more generally in the String Theory framework-- were the SO(10) group occurs naturally and were noncommutativity effects
have been unveiled~\cite{Cecotti:2009zf}.

\section{Acknowledgements}

This work has been financially supported in part by MICINN through grant
FPA2011-24568. I should like to thank P. Schupp for helpful discussions and
encouragement.

\end{document}